\documentclass[aps,prl,groupedaddress,twocolumn]{revtex4}

\usepackage{amsmath,amsfonts,amssymb,graphics,graphicx,epsfig,color,times,bbm,natbib}

\begin{document}

\title{Quantum Repeaters with Photon Pair Sources and Multi-Mode Memories}

\author{Christoph Simon}
\author{Hugues de Riedmatten}
\author{Mikael Afzelius}
\author{Nicolas Sangouard}
\author{Hugo Zbinden}
\author{Nicolas Gisin}
\affiliation{Group of Applied Physics, University of
Geneva, Switzerland}

\date{\today}

\begin{abstract}
We propose a quantum repeater protocol which builds on the
well-known DLCZ protocol [L.M. Duan, M.D. Lukin, J.I.
Cirac, and P. Zoller, Nature {\bf 414}, 413 (2001)], but
which uses photon pair sources in combination with memories
that allow to store a large number of temporal modes. We
suggest to realize such multi-mode memories based on the
principle of photon echo, using solids doped with
rare-earth ions. The use of multi-mode memories promises a
speedup in entanglement generation by several orders of
magnitude and a significant reduction in stability
requirements compared to the DLCZ protocol.
\end{abstract}


\maketitle

The distribution of entanglement over long distances is an
important challenge in quantum information. It would extend
the range for tests of Bell's inequalities, quantum key
distribution and quantum networks. The direct distribution
of entangled states is limited by transmission losses. For
example, 1000 km of standard telecommunications optical
fiber have a transmission of order $10^{-20}$. To
distribute entanglement over such distances, quantum
repeaters \cite{briegel} are likely to be required.
Implementations of quantum repeaters have been proposed in
various systems, including atomic ensembles \cite{DLCZ},
single atoms \cite{Mabuchi}, NV centers \cite{childress}
and quantum dots \cite{VanLoock,simonniquet}. A basic
element of all protocols is the creation of entanglement
between neighboring nodes $A$ and $B$, typically
conditional on the outcome of a measurement, e.g. the
detection of one or more photons at a station between two
nodes. In order to profit from a nested repeater protocol
\cite{briegel}, the entanglement connection operations
creating entanglement between non-neighboring nodes can
only be performed once one knows the relevant measurement
outcomes. This requires a communication time of order
$L_0/c$, where $L_0$ is the distance between $A$ and $B$.
Conventional repeater protocols are limited to a single
entanglement generation attempt per elementary link per
time interval $L_0/c$. Here we propose to overcome this
limitation using a scheme that combines photon pair sources
and memories that can store a large number of
distinguishable temporal modes. We also show that such
memories could be realized based on the principle of photon
echo, using solids doped with rare-earth ions.

Our scheme is inspired by the DLCZ protocol \cite{DLCZ},
which uses Raman transitions in atomic ensembles that lead
to non-classical correlations between atomic excitations
and emitted photons \cite{DLCZexp}. The basic procedure for
entanglement creation between two remote locations $A$ and
$B$ in our protocol requires one memory and one source of
photon pairs at each location, denoted $M_{A(B)}$ and
$S_{A(B)}$ respectively. The two sources are coherently
excited such that each has a small probability $p/2$ of
creating a pair, corresponding to a state
\begin{equation}
\left(1+\sqrt{\frac{p}{2}}(e^{i \phi_A}
a^{\dagger}a'^{\dagger}+e^{i
\phi_B}b^{\dagger}b'^{\dagger})+O(p)\right)|0\rangle.
\label{sources}
\end{equation}
Here $a$ and $a'$ ($b$ and $b'$) are the two modes
corresponding to $S_A$ ($S_B$), $\phi_A$ ($\phi_B$) is the
phase of the pump laser at location $A$ ($B$), and
$|0\rangle$ is the vacuum state. The $O(p)$ term introduces
errors in the protocol, leading to the requirement that $p$
has to be kept small, cf. below. The photons in modes $a$
and $b$ are stored in the local memories $M_A$ and $M_B$.
The modes $a'$ and $b'$ are coupled into fibers and
combined on a beam splitter at a station between $A$ and
$B$. The modes after the beam splitter are
$\tilde{a}=\frac{1}{\sqrt{2}}(a' e^{-i\chi_A} + b'
e^{-i\chi_B}),\tilde{b}=\frac{1}{\sqrt{2}}(a' e^{-i\chi_A}
- b' e^{-i\chi_B})$, where $\chi_{A,B}$ are the phases
acquired by the photons on their way to the central
station. Detection of a single photon in $\tilde{a}$, for
example, creates a state
$|\Phi_{AB}\rangle=\frac{1}{\sqrt{2}}(a^{\dagger} e^{i
\theta_A}+ b^{\dagger} e^{i\theta_B})|0\rangle$ (neglecting
$O(p)$ corrections), with
$\theta_{A(B)}=\phi_{A(B)}+\chi_{A(B)}$, where $a$ and $b$
are now stored in the memories. This can be rewritten as an
entangled state of the two memories,
\begin{equation}
|\Phi_{AB}\rangle=\frac{1}{\sqrt{2}}(|1\rangle_A
|0\rangle_B + e^{i \theta_{AB}}|0\rangle_A|1\rangle_B)
 \label{single},
\end{equation}
where $|0\rangle_{A(B)}$ denotes the empty state of
$M_{A(B)}$, $|1\rangle_{A(B)}$ denotes the state storing a
single photon, and $\theta_{AB}=\theta_B-\theta_A$.

This entanglement can be extended via entanglement swapping
as in Ref. \cite{DLCZ}. Starting from entangled states
$|\Phi_{AB}\rangle=\frac{1}{\sqrt{2}}(a^{\dagger}+e^{i\theta_{AB}}b^{\dagger})|0\rangle$
between memories $M_A$ and $M_B$, and
$|\Phi_{CD}\rangle=\frac{1}{\sqrt{2}}(c^{\dagger}+e^{i\theta_{CD}}d^{\dagger})|0\rangle$
between $M_C$ and $M_D$, one can create an entangled state
between $M_A$ and $M_D$ by converting the memory
excitations of $M_B$ and $M_C$ back into propagating
photonic modes and combining these modes on a beam
splitter. Detection of a single photon after the beam
splitter, e.g. in the mode $\frac{1}{\sqrt{2}}(b+c)$, will
create an entangled state of the same type between $M_A$
and $M_D$, namely
$\frac{1}{\sqrt{2}}(a^{\dagger}+e^{i(\theta_{AB}+\theta_{CD})}d^{\dagger})|0\rangle$.
In this way it is possible to establish entanglement
between more distant memories, which can be used for
quantum communication as follows \cite{DLCZ}.

Suppose that location $A$ ($Z$) contains a pair of memories
$M_{A1}$ and $M_{A2}$ ($M_{Z1}$ and $M_{Z2}$), and that
entanglement has been established between $M_{A1}$ and
$M_{Z1}$, and between $M_{A2}$ and $M_{Z2}$, i.e. that we
have a state $\frac{1}{2}(a_1^{\dagger}+e^{i
\theta_1}z_1^{\dagger})(a_2^{\dagger}+e^{i\theta_2}z_2^{\dagger})|0\rangle$.
The projection of this state onto the subspace with one
memory excitation in each location is
\begin{equation}
|\Psi_{AZ}\rangle=\frac{1}{\sqrt{2}}(a_1^{\dagger}
z_2^{\dagger}+e^{i(\theta_2-\theta_1)}a_2^{\dagger}
z_1^{\dagger})|0\rangle, \label{ent2p} \end{equation} which
is analogous to conventional polarization or time-bin
entangled states. The required projection can be performed
post-selectively by converting the memory excitations back
into photons and counting the number of photons in each
location. Measurements in arbitrary bases are possible by
combining modes $a_1$ and $a_2$ (and also $z_1$ and $z_2$)
on beam splitters with appropriate transmission
coefficients and phases.

The repeater scheme described above is attractive because
it requires only pair sources, photon memories and linear
optical components. The reliance on a single detection for
the elementary entanglement creation makes it less
sensitive to fiber losses than schemes based on coincident
two-photon detections \cite{twophoton}. The price to pay is
the requirement of phase stability, cf. below.

The time required for a successful creation of an entangled
state of the form Eq. (\ref{ent2p}) is given by
\begin{equation}
T_{tot}=\frac{L_0}{c}\frac{1}{P_0 P_1 ... P_n P_{pr}}
\left(\frac{3}{2}\right)^{n+1}. \label{Ttot}
\end{equation}
Here $L_0=L/2^n$, where $L$ is the total distance and $n$
is the nesting level of the repeater. The basic clock
interval is $L_0/c$, the time required for the photons to
propagate from the sources to the central station and for
the information about the result to propagate back to the
memories. The success probability for entanglement creation
for a single elementary link is denoted by $P_0$; $P_i$ is
the success probability for entanglement swapping at the
$i$-th level, and $P_{pr}$ is the probability for a
successful projection onto the state Eq. (\ref{ent2p}). The
probabilities $P_i$ and $P_{pr}$, which are calculated in
Ref. \cite{DLCZ}, depend on the detection and memory
efficiencies. The factors of $\frac{3}{2}$ arise because
entanglement has to be generated for two links before every
entanglement connection. If the average waiting time for
entanglement generation for one link is $T$, there will be
a success for one of the two after $\frac{T}{2}$; then one
has to wait a time $T$ on average for the second one,
giving a total of $ \frac{3T}{2}$.

\begin{figure}
\includegraphics[width=\columnwidth]{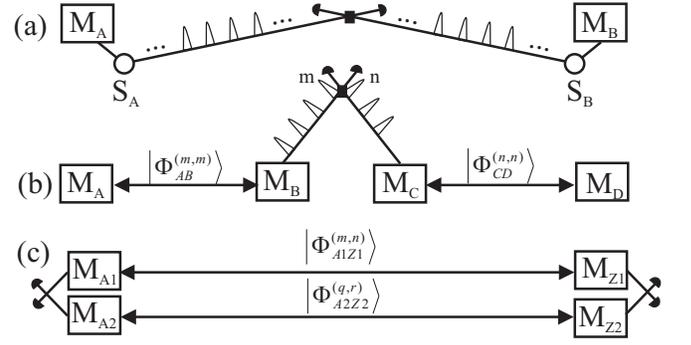}
\caption{Quantum repeater scheme using pair sources and
multi-mode memories. (a) The sources $S_A$ and $S_B$ can
each emit a photon pair into a sequence of time bins. The
detection of a single photon behind the beam splitter at
the central station projects the memories $M_A$ and $M_B$
into an entangled state Eq. (2). (b) If entangled states
have been established between the $m$-th time bins in $M_A$
and $M_B$, and between the $n$-th time-bins in $M_C$ and
$M_D$, an entangled state between the $m$-th time bin in
$M_A$ and the $n$-th time bin in $M_D$ can be created by
re-converting the memory modes into photonic modes and
combining the appropriate time-bins on a beam splitter. (c)
Useful entanglement can be created between two pairs of
distant memories as in Ref. \cite{DLCZ}. Again the
appropriate time bins have to be combined on beam
splitters.} \label{setup}
\end{figure}

For a single creation of the state Eq. (\ref{sources}) per
interval $L_0/c$ the success probability $P_0$ is equal to
$P_0^{(1)}=p \eta_{L_0} \eta_D$, where $\eta_D$ is the
photon detection efficiency and $\eta_{L_0}=e^{-L_0/(2
L_{att})}$, with $L_{att}$ the fiber attenuation length;
$P_0^{(1)}$ is typically very small. However, photon pair
sources can have repetition rates much higher than $c/L_0$,
which is of order 1 kHz for $L_0$ of order 200 km. This
leads one to ask whether it is possible to make several
entanglement creation attempts per interval $L_0/c$. The
source $S_A$ then produces pairs of photons in correlated
pairs of temporal modes (``time bins'') $a_k,a'_k$, with
$k=1,...,N$. All the modes $a_k$ are stored in the memory
$M_A$, and analogously for $S_B$ and $M_B$. If there is a
detection behind the central beam splitter for the $m$-th
time bin, for example in the mode
$\tilde{a}_m=\frac{1}{\sqrt{2}}(a'_m e^{-i\chi_A} + b'_m
e^{-i\chi_B})$, then we know that a state
$|\Phi^{(m,m)}_{AB}\rangle=\frac{1}{\sqrt{2}}(a_m^{\dagger}
e^{i\theta_A}+ b_m^{\dagger} e^{i\theta_B})|0\rangle$ is
stored in the memories $M_A$ and $M_B$, cf. Fig. 1(a).
Running the same protocol for another pair of sources
$S_C,S_D$ and memories $M_C,M_D$, there may be a detection
in the $n$-th time bin, leading to a state
$|\Phi^{(n,n)}_{CD}\rangle=\frac{1}{\sqrt{2}}(c_n^{\dagger}
e^{i\theta_C}+ d_n^{\dagger} e^{i\theta_D})|0\rangle$ being
stored in the memories $M_C$ and $M_D$. One can then
perform entanglement swapping by re-converting the memory
modes $b_m$ and $c_n$ into photonic modes and combining
them on a beam splitter, cf. Fig. 1(b). This leads to an
entangled state
$|\Phi^{(m,n)}_{AD}\rangle=\frac{1}{\sqrt{2}}(a_m^{\dagger}
e^{i(\theta_A+\theta_C)}+ d_n^{\dagger}
e^{i(\theta_B+\theta_D)})|0\rangle$ between the $m$-th mode
stored in $M_A$ and the $n$-th mode stored in $M_D$.
Entanglement of the type Eq. (\ref{ent2p}) can be created
as before, again by combining the appropriate time bins,
cf. Fig. 1(c).

The described protocol requires memories that allow to
store and retrieve the various temporal modes $a_i,b_i$
etc., preserving their distinguishability. We refer to such
memories as {\it multi-mode memories (MMMs)}. We describe
below how MMMs can be realized based on the photon echo
principle, which ensures that photons absorbed at different
times are emitted at different times. With MMMs, working
with $N$ attempts per interval increases the success
probability from $P_0^{(1)}=p\eta_{L_0} \eta_D$ to
$P_0^{(N)}=1-(1-P_0^{(1)})^N$, which is approximately equal
to $N P_0^{(1)}$ for $N P_0^{(1)} \ll 1$, increasing the
overall success rate of the repeater by a factor of $N$.
Our approach based on pair sources and MMMs can be used to
speed up other protocols by the same factor, in particular
schemes based on coincident two-photon detection
\cite{twophoton}, because the speedup occurs at the most
basic level, that of elementary entanglement generation.
There is no obvious equivalent to the use of MMMs as
described above within the Raman-transition based approach
of Ref. \cite{DLCZ}, since all stored modes would be
retrieved at the same time, when the relevant control beam
is turned on. Other forms of multiplexing (spatial,
frequency) can be applied in a similar way both to our
protocol and the DLCZ protocol \cite{collins}.

We now discuss how to realize the elements of our proposal
in practice. Photon pair sources with the required
properties (sufficiently high $p$, appropriate bandwidth,
cf. below) can be realized both with parametric
down-conversion \cite{PDC} and with atomic ensembles
\cite{DLCZexp,atomsources}. Several approaches to the
realization of photon memories have been proposed and
studied experimentally, including EIT \cite{EIT},
off-resonant interactions \cite{offres} and photon echo
\cite{echo}. The echo approach lends itself naturally to
the storage of many temporal modes. Storage and retrieval
of up to 1760-pulse sequences has been demonstrated
\cite{lin}. The temporal information is stored in the
relative phases of atomic excitations at different
frequencies. Photon echoes based on controlled reversible
inhomogeneous broadening (CRIB) \cite{CRIB,hetet} allow in
principle perfect reconstruction of the stored light. The
method is well adapted to atomic ensembles in solids, e.g.
crystals doped with rare-earth ions. To implement such a
memory, one has to prepare a narrow absorption line inside
a wide spectral hole, using optical pumping techniques
\cite{pumping}. The line is artificially inhomogeneously
broadened, e.g. by applying an electric field gradient.
Then the light can be absorbed, e.g. a train of pulses as
described above. After the absorption the electric field is
turned off, and atoms in the excited state are transferred
by a $\pi$-pulse to a second low-lying state, e.g. a
different hyperfine state. For recall, the population is
transferred back to the excited state by a
counter-propagating $\pi$-pulse, and the electric field is
turned back on with the opposite sign, thereby inverting
the inhomogeneous broadening. This leads to a time reversal
of the absorption. The pulse train is re-emitted in
inverted order, with a retrieval efficiency that is not
limited by re-absorption \cite{CRIB}. Photons absorbed in
different memories at different times can be re-emitted
simultaneously (as in Fig. 1) by choosing appropriate times
for the sign-flip of the applied electric field.

The achievable memory efficiency is \cite{sangouard}
\begin{equation}
\eta_M(t)=(1-e^{-\alpha_0 L \gamma_0/\gamma})^2
\mbox{sinc}^2(\gamma_0 t), \label{eta}
\end{equation}
where for simplicity we consider square spectral atomic
distributions, both for the initial narrow line and the
artificially broadened line. Here $\alpha_0 L$ is the
optical depth of the medium before the artificial
broadening, $\gamma_0$ is the initial spectral distribution
width, $\gamma$ is the width after broadening, and $t$ is
the time before transfer to the hyperfine state (neglecting
hyperfine decoherence). The above formula is exact for all
pulse shapes whose spectral support is completely inside
the square atomic distribution. Otherwise there are
additional losses due to spectral truncation of the pulse.
The width $\gamma$ has to be large enough to allow for
pulse durations significantly shorter than the interval
between pulses $\Delta t$, in order to avoid errors due to
pulse overlap. For truncated Gaussian pulses choosing
$\gamma \Delta t=6$ is sufficient for such errors to be
negligible. On the other hand, $\gamma$ is required to be
smaller than the separation between the hyperfine states
used in the memory protocol and whatever state is used for
shelving unwanted atoms in the preparation of the initial
spectral hole (e.g. another hyperfine state). The initial
width $\gamma_0$  has to fulfill $\gamma_0
> 2 \gamma_h$, where $\gamma_h$ is the homogeneous
linewidth of the relevant transition \cite{width};
$\gamma_0$ should be chosen such as to optimize
$\bar{\eta}_M$, the average of Eq. (\ref{eta}) over all
time-bins, i.e. for $t$ between $0$ and $N \Delta t$. One
can show that $\bar{\eta}_M$ can be expressed as a function
of the two variables $x=\gamma_0 N \Delta t$ and
$y=\alpha_0 L/N$.  By adjusting $\gamma_0$ one can choose,
for a given value of $y$, the value of $x$ that maximizes
$\bar{\eta}_M$. Then $\bar{\eta}_M$ becomes a function of
$y$ only, which is plotted in Fig. 2.

\begin{figure}
\includegraphics[width=0.9 \columnwidth]{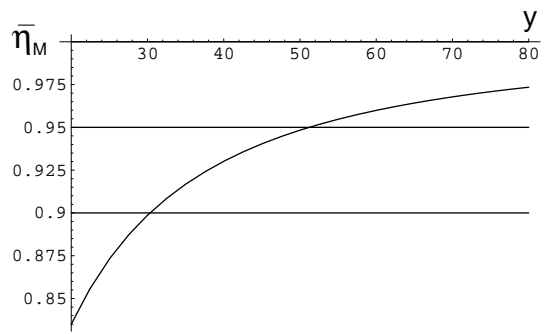}
\caption{The maximum achievable average efficiency
$\bar{\eta}_M$ as a function of $y=\alpha_0 L/N$. One can
see that $\bar{\eta}_M=0.9$ is obtained for $y=30$. This
implies that storing $N=100$ time-bins with this average
efficiency requires $\alpha_0 L=3000$, while 1000 time-bins
require $\alpha_0 L=30000$; $\bar{\eta}_M=0.95$ is obtained
for $y=50$. The required values of $x=\gamma_0 N \Delta t$
are $x=0.8$ for $\bar{\eta}_M=0.9$ and $x=0.6$ for
$\bar{\eta}_M=0.95$.} \label{etaM}
\end{figure}

The storage has to be phase-preserving, so as to conserve
the entanglement for the states of Eqs. (2-3). Decoherence
can affect the excited states during the absorption of the
pulse train, and the hyperfine ground states during the
long-term storage. In rare-earth ions, excited state
coherence times ranging from tens of $\mu$s to 6 ms
\cite{nd-er-paper} and hyperfine coherence times as long as
30 s \cite{hyperfine} have been demonstrated. For a recent
experimental investigation of phase coherence in photon
echo with rare-earth ions see Ref. \cite{staudt}.

We now discuss how to achieve high values for $N$ and
$\bar{\eta}_M$ experimentally. Pr:Y$_2$SiO$_5$ is a very
promising material for initial experiments, since excellent
hyperfine coherence \cite{hyperfine} and memory
efficiencies of order 13 \% \cite{hetet} for macroscopic
light pulses have already been demonstrated. The main
drawback of Pr is the small hyperfine separation (of order
a few MHz), which limits the possible pulse bandwidth and
thus $N$, since by definition $N \leq \frac{L_0}{c\Delta
t}=\frac{L_0 \gamma}{6 c}$ . Neodymium and Erbium have
hyperfine separations of hundreds of MHz
\cite{nd-macfarlane,er-hyperfine}. Nd also has strong
absorption, e.g. Nd:YVO$_4$ with a Nd content of 10 ppm has
an absorption coefficient $\alpha_0=100$/cm
\cite{nd-er-paper} at 879 nm. Choosing $\gamma=300$ MHz
(which gives $\Delta t=6/\gamma=20$ ns) and $\gamma_0=100$
kHz, which is well compatible with $\gamma_h=10$ kHz as
measured for Nd in \cite{nd-er-paper}, our above
calculations show that e.g. $N=400$ and $\bar{\eta}_M=0.9$
would be possible with $\alpha_0 L=30 N=12000$, which could
be achieved with a multi-pass configuration. Erbium-doped
materials can combine good optical coherence and large
inhomogeneous linewidth, e.g. $\gamma_h=2$ kHz and
$\Gamma_{inh}=250$ GHz for Er:LiNbO$_3$ \cite{nd-er-paper},
which makes Er a natural candidate for the implementation
of frequency multiplexing in addition to temporal
multiplexing. The protocol could be run in parallel for a
large number of frequency channels. Even taking into
account the lower absorption for Er, this might allow one
to gain an order of magnitude or more for the overall value
of $N$ compared to our Nd example.

To assess the potential performance of our scheme, consider
a distance $L=1000$ km, and a fiber attenuation of 0.2
dB/km, corresponding to telecom wavelength photons. Note
that the wavelengths of the photon propagating in the fiber
and of the photon stored in the memory can be different in
our scheme. Assume $\bar{\eta}_M$=0.9 and
photon-number-resolving detectors with efficiency
$\eta_D=0.9$. Highly efficient number-resolving detectors
are being developed \cite{kim,supercon}. One can show that
the optimal nesting level for the repeater protocol for
these values is $n=2$, corresponding to $2^n=4$ elementary
links, which gives $L_0=250$ km. Using Eq. (\ref{Ttot}) and
Ref. \cite{DLCZ} one can show that the total time for
creating a state of the form Eq. (3) using the scheme of
Fig. 1 is
\begin{equation}
T_{tot}=\frac{L_0}{c}\frac{18 (2-\eta)(4-3\eta)}{N p\,
\eta_{L_0}\eta_D\eta^4},
\end{equation}
where $\eta=\bar{\eta}_M \eta_D$ and $\eta_{L_0}=e^{-L_0/(2
L_{att})}$, with $L_{att}=22$ km, and $c=2\times 10^8$ m/s
in the fiber. One can show by explicit calculation of the
errors due to double emissions \cite{simon} that the
fidelity $F$ of the final entangled state compared to the
ideal maximally entangled state for a repeater with $n=2$
levels is approximately $F=1-56 (1-\eta) p$. If one wants,
for example, $F=0.9$ one therefore has to choose $p=0.009$,
which finally gives $T_{tot}=3400/N$ s. If one can achieve
$\bar{\eta}_M=\eta_D=0.95$ one finds $T_{tot}=800/N$ s.
High-efficiency MMMs as discussed above could thus reduce
$T_{tot}$ for 1000 km to a few seconds (or less with
frequency multiplexing).

MMMs can also help to significantly alleviate the stability
requirements for the repeater protocol. For simplicity, let
us just consider an elementary link (between locations $A$
and $B$) from our above example, with $L_0=250$ km. The
entanglement in Eq. (\ref{ent2p}) depends on the phase
difference $\theta_2-\theta_1$, which can be rewritten as
$(\theta_{B_2}(t_2)-\theta_{B_1}(t_1))-(\theta_{A_2}(t_2)-\theta_{A_1}(t_1))$.
Here $t_1$ ($t_2$) is the time when the first (second)
entangled state of the type of Eq. (\ref{single}) is
created. The phases thus have to remain very stable on the
timescale given by the typical value of $t_2-t_1$
\cite{fibers}. In the case without MMMs the mean value of
$t_2-t_1$ is $L_0/(cP_0)$, which is of order 10 s for our
above example ($L_0/c$ of order 1 ms and $P_0$ of order
$10^{-4}$). Over such long timescales, both the phases of
the pump lasers and the fiber lengths are expected to
fluctuate significantly. Active stabilization would thus
definitely be required. With MMMs, for large values of $N$,
$P_0$ can be made sufficiently large that it becomes
realistic to work only with states Eq. (\ref{ent2p}) where
the initial entanglement between $A_1$ and $B_1$ and
between $A_2$ and $B_2$ was created in the same interval
$L_0/c$ (the probability for such a double success is
$P_0^2$). For our above example, $P_0$ can be made of order
$10^{-1}$ for $N$ of order $10^3$. Working only with
entangled states from the same interval increases the time
$T_{tot}$ by a factor of order $1/P_0$, but it reduces the
mean value of $t_2-t_1$ to of order $N\Delta t$, which is
of order 20 $\mu$s for our example. For such short time
scales, active stabilization of the laser and fiber phases
may not be required.

In conclusion the combination of photon pair sources and
multi-mode memories should allow the realization of a
quantum repeater protocol that is much faster and more
robust than the protocol of Ref. \cite{DLCZ} while
retaining its attractive features, in particular the use of
linear optical elements and of single photon detections for
entanglement generation and swapping.

{\it Acknowledgements.} We thank M. Halder, S. Kr\"{o}ll,
J. Min\'{a}\v{r}, V. Scarani, and W. Tittel for useful
discussions. This work was supported by the EU Integrated
Project {\it Qubit Applications} and by the Swiss NCCR {\it
Quantum Photonics}.

\bibliographystyle{apsrev}

\begin{thebibliography}{31}
\bibitem{fibers} Note
that it is possible to use the same fibers to connect $A_1$
and $A_2$ and respectively $B_1$ and $B_2$ to the central
station \cite{DLCZ}.
\end{thebibliography}

\end{document}